\documentclass[useAMS,usenatbib]{mn2e}
\usepackage{graphicx}
\begin{document}
\newcommand{\comment}[1]{{\ }}
\newcommand{\degree}{^{\circ}}
\def\Msun{\ifmmode{~{\rm M}_\odot}\else${\rm M}_\odot$~\fi}
\newcommand{\der}[2]{{\frac{d#1}{d#2}}}
\newcommand{\doo}[2]{{\frac{\partial#1}{\partial#2}}}
\newcommand{\fat}[1]{{\bmath{ #1}}}
\newcommand{\FAT}[1]{{\mathbfss{ #1}}} 
\newcommand{\half}{{\frac12}}
\title[Regularization with velocity-dependent forces]{Algorithmic regularization with velocity-dependent forces}

\author[S.\ Mikkola and D.\ Merritt]
{
Seppo Mikkola$^1$\thanks{E-mail: Seppo.Mikkola@utu.fi} and 
David Merritt$^2$\thanks{E-mail: David.Merritt@rit.edu}
\newauthor
$^1$Tuorla Observatory, University of Turku, 
V\"ais\"al\"antie 20, Piikki\"o, Finland
\newauthor
$^2$Department of Physics, Rochester Institute of Technology, 
Rochester,NY 14623, USA
}

\date{\today}

\pagerange{\pageref{firstpage}--\pageref{lastpage}} \pubyear{}

\maketitle

\label{firstpage}

\begin{abstract}
Algorithmic regularization  uses a transformation of the
equations of motion such that the leapfrog algorithm produces
exact trajectories for two-body motion as well as
regular results in numerical integration of the motion of strongly
interacting few-body systems.
That algorithm alone is not sufficiently 
accurate and one must use the extrapolation method
for improved precision. This requires that the basic
leapfrog algorithm be time-symmetric, which is not directly possible 
in the case of velocity-dependent forces, but 
is usually obtained with the help of the implicit midpoint method. 
Here we suggest an alternative explicit algorithmic 
regularization algorithm which can handle velocity-dependent forces.
This is done with the help of a generalized
midpoint method to obtain the required time symmetry, 
thus eliminating the need for the implicit midpoint
method and allowing the use of extrapolation.    
\end{abstract}


\section{Introduction}

In some $N$-body problems one has velocity-dependent 
perturbations. Examples are the relativistic terms, 
which are important in black hole dynamics \citep{Aarseth2003}, 
or dissipative terms due to tidal friction or 
atmospheric friction in satellite orbits. 
The KS-regularization (e.g. basic KS: \citealt{KS,LinReg}
and the {\sevensize CHAIN}-method of \citealt{CHAIN})
can easily handle any additional forces, 
however in multi-body regularization with the KS-transformation,
large mass ratios cause problems. 
Therefore other regularization methods
 --algorithmic regularizations--
such as the logarithmic Hamiltonian method
\citep{Algoreg,logH,PretoTremaine1999} 
or the time-transformed leapfrog \citep{TTL} must be considered. 
On the other hand, these methods, 
when combined with the extrapolation method
\citep{Gragg1964,Gragg1965,BS}
cannot easily include velocity-dependent forces, 
except with the help of the implicit midpoint method. 
Since implicit methods may be inefficient, there is motivation
to study ways to make the integrations explicit, 
while at the same time utilizing the good properties 
of algorithmic regularization. 

Algorithmic regularization is simpler than KS regularization and, 
what is most important,
versions of it work for arbitrary mass ratios. 
This is especially important in simulations of black hole dynamics 
in galactic nuclei \citep{Merritt2006}.

In this paper, we first introduce the problem using a perturbed
two-body system as an example. 
Then we suggest a generalized midpoint method
to be used as a tool to time-symmetrize any basic algorithm.
Finally the generalization to the $N$-body problem is briefly outlined. 

\section{Generalized Algorithmic Regularization}

Here we discuss the formulation of the basic algorithms, 
the time-transformed leapfrogs, that are regular in two-body 
collisions.
Then a generalized midpoint method, that
can also be used with the 
Bulirsch-Stoer (BS) extrapolation method 
\citep{Gragg1964,Gragg1965,BS},
is introduced.

\subsection{The perturbed two-body problem}
We first consider the perturbed two-body problem with 
velocity-dependent forces.
Let $\fat r $ and $\fat v$ be the position and velocity vectors respectively
and $m$ the mass of the two-body system and $t$ the time. 
We may then write the equation of motion as
\begin{eqnarray}
\label{twobodyequations}
\dot \fat v&=&-m\frac{\fat r}{r^3}+\fat f(\fat r,t,\fat v), \\
\dot \fat r&=&\fat v.
\end{eqnarray}
This case is simple enough for a detailed discussion; 
generalization to the full $N$-body problem will be straightforward.

As \cite{Algoreg,logH} and \cite{PretoTremaine1999} demonstrated,
there is a way to make the leapfrog algorithm exact for two-body orbits, 
and regular for two-body collisions in more complicated problems, 
if one introduces a time transformation. 
Here we concisely re-derive the algorithm and augment it
to the case of a general (not necessarily Hamiltonian) perturbation.

Let
\begin{equation}
 b=\frac{m}{r}-\half \fat v^2
\end{equation}
 be the binding (Kepler) energy of the two-body system. 
We have  the energy equations
\begin{eqnarray}
\half\fat v^2+b = \frac{m}{r}, \\
\dot b = -\fat v\cdot \fat f.
\end{eqnarray}
This allows the introduction of the two time transformations
\begin{eqnarray}
\frac{dt}{ds} &=&\frac1{\half\fat v^2+b} \label{tpr1}, \\
\frac{dt}{ds} &=& \frac{r}{m},   \label{tpr2}
\end{eqnarray}
which are equivalent along the solution trajectory.
Using the first alternative (\ref{tpr1}) to 
transform the equation of motion for
the coordinates $(t,\fat r)$,
one gets
\begin{eqnarray}\label{ttlequations}
t'&=&\frac{1}{\half \fat v^2+b}, \\
 \fat r'&=&\frac{\fat v}{\half \fat v^2+b},
\end{eqnarray}
and the second equation gives for $b$ and $\fat v$ 
\begin{eqnarray}
b'&=&-\fat v\cdot \fat g, \\
  \fat v'&=&-\frac{\fat r}{r^2}+\fat g ,
\end{eqnarray}
where primes indicate differentiation with respect to the new 
independent variable $s$ and  
\begin{equation}\fat g=\frac{r}{m}\fat f(\fat r,t,\fat v).\end{equation}
If the perturbation $\fat f $ (hence $\fat g$) is independent of the
velocity $\fat v$, 
then the above equations allow the use of the leapfrog algorithm:
\begin{eqnarray}\label{lfbegin}
t_\half&=&t_0+\frac{h}2 \frac{1}{\half \fat v_0^2+b_0}, \\
\fat r_\half&=&\fat r_0+\frac{h}{2}\frac{\fat v_0}{\half \fat v_0^2+b_0}, \\
\label{v1formula1}
\fat v_1&=&\fat v_0-h\frac{\fat r_\half}{r_\half^2}+h\fat g_\half, \\
b_1&=&b_0-h \fat v_\half \cdot \fat g_\half, \\
t_1&=&t_\half+\frac{h}2 \frac{1}{\half \fat v_1^2+b}, \\
\fat r_1&=&\fat r_\half+\frac{h}{2}\frac{\fat v_1}{\half \fat v_1^2+b_1},
\label{lfend}
\end{eqnarray}
where the subscripts 0 and 1 refer to the beginning and the end of the step,
and $\fat v_\half=(\fat v_0+\fat v_1)/2$.
If the perturbation $\fat g=\fat 0$, then the motion is pure Kepler motion
and the leapfrog algorithm produces an exact trajectory with only
a time error \citep{Algoreg,logH,PretoTremaine1999}.

In the above equations, the symbol $\fat g_\half$ indicates
$\fat g(\fat r_\half)$.
However, if $\fat g$ actually depends on the velocity too, then
the leapfrog cannot be immediately formed. 
This problem (or rather an analogous one) was solved by \cite{TTL}, 
using the implicit midpoint method,
i.e. it was necessary to solve the equation
\begin{equation}
\label{v1formula2}
\fat v_1=\fat v_0-h\frac{\fat r_\half}{r_\half^2}+h\fat 
g\left(\fat r_\half,t_\half,\frac{\fat v_0+\fat v_1}{2}\right)
\end{equation} 
for $\fat v_1$. Often this solution is possible only by iteration which can be
rather expensive if the perturbation is strong and complicated.
This fact motivates a search for ways to 
find an alternative that is explicit, yet
capable of utilizing the algorithmic regularization. 
This goal can be achieved with the help of the algorithm we next discuss.

\subsection{Generalized midpoint method}
Here we introduce a generalization to the well-known modified
midpoint method. 
In this algorithm, the basic approximation to
advance the solution is not just the evaluation of the derivative
at the midpoints, but any method to approximate the solution.
Thus the algorithmic regularization by the leapfrog can be
used even when the additional force depends on velocities.
That provides a regular basic algorithm, which is made
suitable for the extrapolation method by means of the 
generalized midpoint method, as follows.

Consider the differential equation
\begin{equation}
\label{zequ}
\dot \fat z=\fat f(\fat z),\ \ \fat z(0)=\fat z_0.
\end{equation}
Splitting the above as
\begin{eqnarray}
\dot \fat x&=&\fat f(\fat y), \\
\dot\fat y&=&\fat f(\fat x)
\end{eqnarray}
with the initial values
\[
\fat x_0=\fat y_0=\fat z(0),
\]
gives the leapfrog-like algorithm
\begin{eqnarray}
\fat x_\half&=&\fat x_0+\frac{h}2 \fat f(\fat y_0), \\
\fat y_1&=&\fat y_0+h f(\fat x_\half), \\
\fat x_1&=&\fat x_\half+\frac{h}2 \fat f(\fat y_1).
\end{eqnarray}
However, this is nothing but another way to write the well-known
modified midpoint method.

A new interpretation of the above can be obtained by first
rewriting it in the form
\begin{eqnarray}
\label{yks}\fat x_\half&=&\fat x_0+\left(+\frac{h}2 \fat f(\fat y_0)\right),\\
\label{kaks}\fat y_\half&=&\fat y_0-\left(-\frac{h}2 f(\fat x_\half)\right),\\
\label{kol}\fat y_1&=&\fat y_\half+\left(+\frac{h}2 f(\fat x_\half)\right),\\
\label{nel}\fat x_1&=&\fat x_\half-\left(-\frac{h}2 \fat f(\fat y_1)\right).
\end{eqnarray}
In (\ref{yks}) the bracketed term is an (Euler-method) approximation
to the increment of $\fat x$ over the time interval $h/2$ with the
initial value $\fat y_0$, while in (\ref{kaks}) the initial
value is $\fat x_\half\approx\fat x(h/2)$ and the time interval is $-h/2$
Finally, this increment is added --with a minus sign-- to $\fat y_0$ 
to obtain an approximation for $\fat y(h/2)$. 
In the remaining formulae (\ref{kol}), (\ref{nel}), 
the idea is the same but the roles of $\fat x$ and $\fat y$ 
have been changed.

A generalization of this is now obvious.
Let
\begin{equation}
\fat z(\Delta t)\approx \fat z_0+\fat d(\fat z_0,\Delta t)
\end{equation}
be an  approximation to the solution of Eq. (\ref{zequ})
over a time interval $\Delta t$. 
In Euler's method,
\begin{equation}
d(\fat z_0,\Delta t)=\Delta t \fat f(\fat z_0),
\end{equation}
which gives the algorithm described in Eqs. (\ref{yks}) -- (\ref{nel}), 
but in general,
$\fat d$ could be obtained from any reasonable method for solving
the differential equation (\ref{zequ}). 
We thus choose a method and define
\begin{equation} 
\fat d(\fat z_0,\Delta t)=\widetilde \fat z(\Delta t) -\fat z_0,
\end{equation}
where $\widetilde \fat z(\Delta t)$ is the approximation for 
$\fat z(\Delta t)$
obtained with the chosen method.
This generalized midpoint algorithm may be especially useful 
if one uses a special method that is well-suited to the 
particular problem at hand.

One step in the generalized midpoint method can now be written
\begin{eqnarray}
\label{yksd}\fat x_\half&=&\fat x_0+\fat d(\fat y_0,+\frac{h}2), \\
\label{kaksd}\fat y_\half&=&\fat y_0-\fat d(\fat x_\half,-\frac{h}2), \\
\label{kold}\fat y_1&=&\fat y_\half+\fat d(\fat x_\half,+\frac{h}2), \\
\label{neld}\fat x_1&=&\fat x_\half-\fat d(\fat y_1,-\frac{h}2),
\end{eqnarray}
or, if we define the mapping (or ``subroutine'' )
\begin{eqnarray}\label{AAAsubru1}
\FAT A(\fat x,\fat y,h):\ \ \ \ \label{oned}\fat x&\rightarrow&\fat x+\fat d(\fat y,+\frac{h}2)\\
\label{twod}\fat y&\rightarrow&\fat y-\fat d(\fat x,-\frac{h}2),
\label{AAAsubru2}\end{eqnarray}
we can write the algorithm with many ($N$) steps as
\begin{eqnarray}\nonumber
1.&&{\rm Set\ } \fat y=\fat x; \\ \label{AAA}
2.&&{\rm Repeat \ }\FAT A(\fat x,\fat y,h)\FAT A(\fat y,\fat x,h)\ \ N\ {\rm times};
\label{AAAalgo}\\   \nonumber
3.&&{\rm Accept\  }\fat x {\rm \ as \  the \ final \ result}.
\end{eqnarray}
Thus one simply calls the subroutine $\FAT A $ alternately with
arguments $(\fat x,\fat y)$ and $(\fat y,\fat x)$ such that the sequence 
is time-symmetric (starts and stops with $\fat x$ in Eq. \ref{AAA}).

This basic algorithm has the correct symmetry -- because it was derived 
from a leapfrog-like treatment -- such that the error in integration
over a fixed time interval with different timesteps $h$ can be written
\begin{equation}
{\rm error}=A_1 h^2+A_4 h^4 +..,
\end{equation} 
and thus the Gragg-Bulirsch-Stoer extrapolation method can be used
to obtain high accuracy.

The great advantage of this generalized midpoint method is
that the leapfrog with the implicit midpoint method can be replaced by
a method that is not exactly time-symmetric. 
The computation of the quantity $\fat g_\half$, 
when it depends on velocity, can be done
in a straightforward way, e.g. by
\begin{equation}
\fat g_\half=\fat g(\fat r_\half,t_\half,\fat v_\half),
\end{equation}
where one may approximate $\fat v_\half$ either 
by $\fat v_\half\approx\fat v_0$ or preferably by
\begin{equation}
\fat v_\half\approx\fat v_0-\frac{h}{2}\frac{\fat r_\half}{r_\half^2}
\end{equation}
after which
\begin{equation}
\fat v_1=\fat v_0-h\frac{\fat r_\half}{r_\half^2}+
h\fat g(\fat r_\half,t_\half, v_\half)
\end{equation}
can be used instead of (\ref{v1formula1}) (or \ref{v1formula2}).
Here it is necessary to stress that only the increments of the
variables from the
algorithm (\ref{lfbegin})--(\ref{lfend}) are to be used
as the quantities $\fat d$ in the algorithm 
(\ref{AAAsubru1})--(\ref{AAAsubru2}).

\begin{figure}
\centerline{\includegraphics[angle=0,width=9cm,height=7cm]{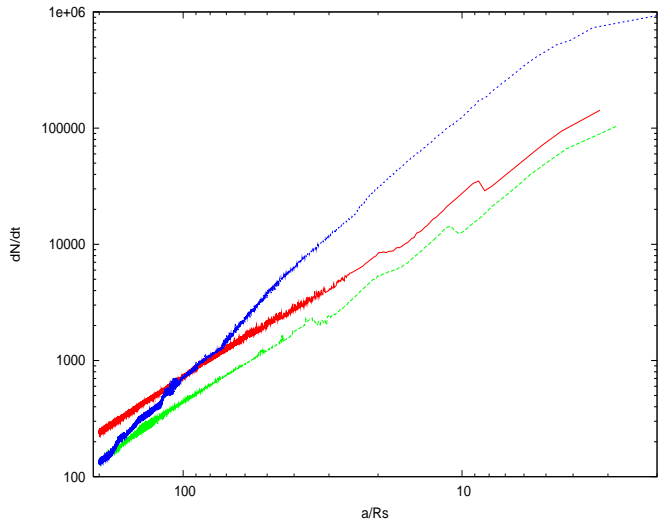}}
\caption{ Number of perturbation evaluations per unit of time
($dN/dt$) in a two-black-hole system,
with masses and speed of light  $m_1=0.9,\  m_2=0.1,\  c=20$,
integrated from the initial values $ a_0=1,\  e_0= 0$ until the final merger
of the two black holes.
The $x$-axis
is the semi-major axis in units of the Schwarzschild radius.
The green curve is for the new method while red and blue illustrate
two varieties of the implicit midpoint method (as described in the
text). 
}
\label{effigu}
\end{figure}

\begin{figure}
\centerline{\includegraphics[angle=0,width=9cm,height=7cm]{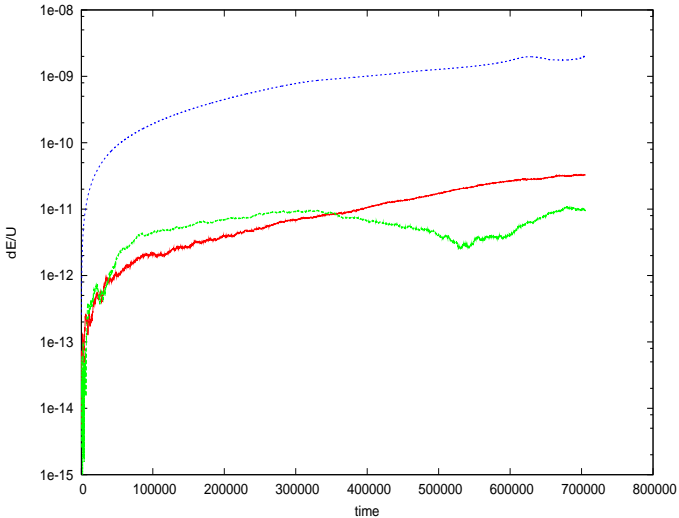}}
\caption{Error evolution in the experiments. 
The measure of error, plotted as a function of time,
is $|(\half\fat v^2+b)\frac{r}{m}-1|$.
Colour coding is the same as in the previous figure.
}
\label{errfigu}
\end{figure}

\section{Some Experiments}\label{experimentsection}
Using a simple perturbed two-body code, written according
to the above theory, we carried out
some experiments to compare the new alternative with 
the implicit midpoint method.

Tests with an (initially) circular orbit of unit radius 
and with the perturbing (frictional) force $\fat f=-\epsilon\fat v$
suggest that for very small $\epsilon$ ( $\le 10^{-6}$) the 
implicit midpoint method is faster, but for stronger perturbations,
the new method is favorable.

Tests with the relativistic PPN2.5 terms from \cite{Soffel1989}
are illustrated in Fig.\ \ref{effigu}. 
Here the system was a two-body system with masses $m_1=0.9,\ m_2=0.1$, 
initial semi-major-axis $a_0=1$, initial eccentricity $e_0=0$ and the 
velocity of light was set to $c=20$.
Due to the gravitational radiation term, the semi-major axis shrinks
and the computational effort ($dN/dt$= number of perturbation evaluations
per unit of time) increases. 
The figure illustrates the evolution of $dN/dt$ 
(averages over $100$ steps with BS extrapolation) 
during the computation (until final merger of the bodies)
for three different methods.  
The results are plotted as a function of the shrinking semi-major axis 
(measured in terms of the Schwarzschild radius for the combined mass).
In these integrations the one-step relative error tolerance was set to 
$10^{-13}$ and the errors, measured via the quantity
$\frac{r}{m}(\half \fat v^2+b)-1$,
were  $\sim 10^{-11}$ for the new method and the midpoint method with 
iteration to convergence (corresponding to the ``green'' and ``red'' 
experiments in Fig.~\ref{effigu}).
 For the restricted iteration method the error was, however, about 
$10^{-9}$ suggesting that this method is not to be recommended. 
The errors grew secularly, as can be seen from Fig.~\ref{errfigu};
the numbers given above refer to the values just at merger,
i.e. when the  two particles approach more closely than the sum of 
Schwarzschild radii.
It may be seen that in all cases, the new method is somewhat more efficient. 

\section{$N$-body formulation}
The generalization of the algorithm to the $N$-body problem
is simple in principle. 
One may use the leapfrog algorithms introduced by 
\cite{Algoreg,logH} or \cite{TTL}
and simply add the necessary velocity-dependent forces.
A new formulation that effectively unifies the above cited
works may be constructed as follows.
 Let $T=(1/2)\sum_k m_k \fat v_k^2$ be the kinetic energy, 
$U=\sum_{i<j}m_im_j|\fat r_i-\fat r_j|^{-1}$ be the potential energy, 
and $\Omega$ an (in principle) arbitrary function of
of the coordinates, often 
\begin{equation}\label{Omegaequation}
\Omega=\sum_{i<j} |\fat r_i-\fat r_j|^{-1}.
\end{equation}
 Then one may define, in analogy with (\ref{tpr1}) and (\ref{tpr2}),
the two time transformations
\begin{equation}
t'=1/(\alpha T+B)=1/(\alpha U+\beta \Omega+\gamma),
\end{equation}
where $\alpha,\ \beta$ and $\gamma$ are adjustable constants.
Since $T=U+E$, we have $B=-\alpha E+\beta \Omega+\gamma$, which expression
is used only for the initial value of $B$ 
and later this quantity  must be obtained by solving
the differential equation
\begin{equation}
\dot B=-\alpha \sum_k \fat v_k\cdot \fat f_k+\beta\sum_k \doo{\Omega}{\fat r_k}\cdot \fat v_k. 
\end{equation}
In the above, $\fat v_k,\ \fat r_k$ are the velocity and position of the
body with mass $m_k$, correspondingly, and the forces additional to
$\partial U/\partial \fat r_k$ are denoted by $\fat f_k$.

The equations of motion that can be used to construct the
leapfrog that provides algorithmic regularization are, 
for time and coordinates respectively,
\begin{eqnarray}
t'&=&1/(\alpha T+B), \\
\fat r_k'&=&t'\fat v_k
\end{eqnarray}
and for velocities and $B$
\begin{eqnarray}
\tau'&=&1/(\alpha U+\beta \Omega+\gamma), \\
\fat v_k'&=&\tau'(\doo{U}{\fat r_k}+\fat f_k)/m_k, \\
B'&=&\tau'\sum_k\left(-\alpha \fat f_k+\beta\doo{\Omega}{\fat r_k}\right)\cdot \fat v_k.
\end{eqnarray}
Here the (possible) velocity dependence of the additional forces $\fat f_k$
can be handled as in our two-body example above. However, to account 
for the (explicitely written)  $\fat v$-dependence of $B'$ one must
follow \cite{TTL}, i.e. first the $\fat v_k$ are advanced and then the
average $(\fat v_k(0)+\fat v_k(h))/2$ is used to evaluate $B'$. 
Thus the leapfrog can be constructed in obvious analogy with
the perturbed two-body case. 
However, in $N$-body integrations,
the roundoff can be a serious source of error and relative coordinates
of close bodies must be used to reduce that effect
\citep{Algoreg,TTL}. 
\smallskip

Some additional remarks follow.

\begin{description}
\item[(i)] If one takes $(\alpha,\beta,\gamma)=(1,0,0)$ then the method 
obtained is the logarithmic Hamiltonian method \citep{Algoreg}.

\item[(ii)] If $(\alpha,\beta,\gamma)=(0,1,0)$ then we have the time 
transformed leapfrog (TTL) \citep{TTL}.

\item[(iii)] If $(\alpha,\beta,\gamma)=(0,0,1)$ then the method is just the
normal basic leapfrog.

\item[(iv)] If there are no velocity-dependent perturbations, then the 
normal leapfrog can be used and it is in fact faster. 
This is because
our alternative algorithm then does some (unnecessary) 
calculations back and forth.

\item[(v)] The question of which combination of the numbers  
$(\alpha,\beta,\gamma)$ is best cannot be answered in general, 
but experimentation is necessary. For $N$-body systems with
very large mass ratios, however, it seems that one must have $\beta\ne 0$,
which means a form of the TTL method.

\item[(vi)] The experiments discussed in section \ref{experimentsection}
correspond to the alternative (i), i.e. $(\alpha,\ \beta,\ \gamma)=(1,0,0)$.
Note that for the case of only two bodies,
there should be not much difference
between alternatives (i) and (ii) since in this case 
they are mathematically equivalent \citep{TTL}.
\end{description}

\section{Conclusion}

We have demonstrated that the generalized midpoint algorithm
can be used to time-symmetrize  the algorithmic regularization leapfrog even
when the forces depend on velocities. 
This permits efficient use of the extrapolation method. 
For very small perturbations, the implicit midpoint method may
still be better, and the new method can be recommended 
only when the velocity dependence of the forces is significant.
Finally we note that the generalized midpoint method
can be used with any special low order approximation to the
differential equations under consideration.
Thus it is not restricted to $N$-body problems.

\label{lastpage}
\end{document}